\documentclass{article}

\begin{document}

\markboth{P.G. Estevez, M.L. Gandarias and J. Lucas} {Classical Lie symmetries and reductions of a
nonisospectral Lax pair}

\title{Classical Lie symmetries and reductions of a nonisospectral Lax pair}

\author{ P.G. ESTEVEZ $^{a}$, M. L. GANDARIAS $^b$ and J. LUCAS $^c$ \\
$^{a}$ Departamento de Fisica Fundamental, Universidad de Salamanca,  SPAIN\\
email: {pilar@usal.es}\\
$^b$ Departamento de Matematicas, Universidad de Cadiz, SPAIN\\ $^c$ Institute of Mathematics, Polish Academy of Sciences,
Warszawa, POLAND}

\maketitle

\begin{abstract}
The classical Lie method is applied to a nonisospectral problem associated with a  system of partial
differential equations in $2+1$ dimensions (Maccari A, {\it J. Math. Phys} {\bf 39}, (1998), 6547-6551).
Identification of the classical Lie symmetries provides a set of  reductions that give rise to different
nontrivial spectral problems in $1+1$ dimensions. The form in which the spectral parameter of the $1+1$ Lax
pair is introduced is carefully described.
\end{abstract}



Accepted in Journal of Nonlinear Mathematical Physics

\section{Introduction}
Some of the authors \cite{egp05}  have applied the classical Lie Method \cite{bluco}, \cite{olver} not just
to a partial differential equation (PDE) in $2+1$ dimensions, but to the Lax pair associated with this PDE
\cite{legar96}. This requires consideration not only of the independent variables and the fields involved in
the PDE, but also the eigenfunctions of the Lax pair. The bonus is that once we have identified the
symmetries, we can proceed to the corresponding $1+1$ reductions \cite{stephani}, which provide us with the
reduced $1+1$ equations as well as their associated spectral problems. The spectral parameter of the reduced
$1+1$ Lax pair is introduced in a very natural way.

At this point, it is interesting to recall the  Ablowitz-Ramani-Segur conjecture \cite{ars}, which
establishes that a PDE is integrable in the Painlev\'e sense \cite {pp} if all its reductions pass the
Painlev\'e test \cite{Weiss}. This means that solutions of a PDE can be achieved by solving its reductions to
ordinary differential equations (ODE). Nevertheless, it is well known \cite{ep01} that  it is often more
difficult to find  the linear problem associated with a $1+1$ PDE than with a $2+1$ PDE. In this sense, our
approach is  the opposite of that of  Ablowitz-Ramani-Segur. We start with a $2+1$ spectral problem and we
obtain several nontrivial $1+1$ Lax pairs through classical Lie reductions of the former.

The problem  we are considering here concerns  the following $2+1$ system of PDEs:
\begin{eqnarray}
0&=& u_t+u_{xxx}+u_{xy}-6u\omega u_x +2m_yu\label{1.1}\\
 0&=& \omega_t+\omega_{xxx}-\omega_{xy}-6u\omega \omega_x
-2m_y\omega\label{1.2}\\
0&=& m_x+u\omega. \label{1.3}
\end{eqnarray}

This system is the real version  of the PDE proposed by Maccari in \cite{Maccari98}. This equation is a
particular member of a class of integrable equations found by Calogero and Degasperis in \cite{calogero} and
constitutes a $2+1$  generalization of the equation proposed by Hirota in \cite{Hirota}. In reference
\cite{estevez01} one of us proved that the system has the Painlev\'e property and the singular manifold
method \cite {Weiss} was used to derive
 the following nonisospectral
two-component Lax pair associated with the system (\ref{1.1})-(\ref{1.3}).
\begin{eqnarray}
0&=&\psi_x+\frac{\lambda}{2}\psi+u\chi \label{1.4}\\
0&=&\chi_x-\frac{\lambda}{2}\chi+\omega\psi \label{1.5} \\
0&=&\psi_t-\lambda\psi_y+\left[m_y-\omega u_x+u\omega_x+\lambda u
\omega-\frac{1}{2}\lambda_y-\frac{1}{2}\lambda^3\right]\psi+\nonumber\\ &+&\left[2\omega
u^2-u_y-u_{xx}+\lambda u_x-\lambda^2 u\right]\chi \label{1.6}\\0 &=&\chi_t-\lambda\chi_y+\left[-m_y+\omega
u_x-u\omega_x-\lambda u \omega-\frac{1}{2}\lambda_y+\frac{1}{2}\lambda^3\right]\chi+\nonumber\\
&+&\left[2\omega^2 u+\omega_y-\omega_{xx}-\lambda \omega_x-\lambda^2 \omega\right]\psi.\label{1.7}
\end{eqnarray}

The system is particularly interesting because the compatibility conditions
$\psi_{xt}=\psi_{tx},\chi_{xt}=\chi_{tx}$ imply that the spectral parameter $\lambda$ is a function of $y$
and $t$ that satisfies:
\begin{equation}\lambda_t-\lambda\lambda_y=0,\qquad \lambda_x=0.\label{1.8}\end{equation}

The possible $1+1$ dimensional reductions of (\ref{1.1})-(\ref{1.3}) are of course interesting, but it is
better to study the reductions of the spectral problem (\ref{1.4})-(\ref{1.7}) because, by doing this, we
shall know how the spectral parameter appears in the reduction and this is a nontrivial question at all as we
shall see.

In section (2), we apply  the Classical Lie Method of finding point symmetries to (\ref{1.4})-(\ref{1.7})
system. The different possible reductions corresponding to these symmetries are identified in section 3.
Different non trivial $1+1$ spectral problems are obtained through these reductions. The conclusions are
presented in section 4.

\section{Classical Symmetries}
 In order to apply the Classical Lie Method \cite{stephani} to the system of PDEs (\ref{1.4})-(\ref{1.7}) with
three independent variables and six fields, we consider the one-parameter Lie group of infinitesimal
transformations  given by:
\begin{eqnarray}
 &&x' = x + \varepsilon\,\xi_1(x,y,t,u,\omega,m,\psi,\chi,\lambda) + O(\varepsilon^2)\\&& y' =y +
\varepsilon\,\xi_2(x,y,t,u,\omega,m,\psi,\chi,\lambda) + O(\varepsilon^2)\\
 &&t'  = t +\varepsilon\,\xi_3(x,y,t,u,\omega,m,\psi,\chi,\lambda) + O(\varepsilon^2),
\end{eqnarray}

and
\begin{eqnarray}
 &&\nonumber u'  = u + \varepsilon\,\phi_1(x,y,t,u,\omega,m,\psi,\chi,\lambda) + O(\varepsilon^2)\\&&
\omega'  = \omega + \varepsilon\,\phi_2(x,y,t,u,\omega,m,\psi,\chi,\lambda) + O(\varepsilon^2)\\ &&
 m' = m + \varepsilon\,\phi_3(x,y,t,u,\omega,m,\psi,\chi,\lambda) + O(\varepsilon^2)\\ &&
\psi'  = \psi + \varepsilon\,\phi_4(x,y,t,u,\omega,m,\psi,\chi,\lambda) + O(\varepsilon^2)\\ &&
 \chi'  =
\chi + \varepsilon\,\phi_5(x,y,t,u,\omega,m,\psi,\chi,\lambda) + O(\varepsilon^2)\\&& \lambda'  = \lambda +
\varepsilon\,\phi_6(x,y,t,u,\omega,m,\psi,\chi,\lambda) + O(\varepsilon^2) , \end{eqnarray} where $\epsilon$
is the group parameter. This transformation must therefore leave the set of solutions of
(\ref{1.4})-(\ref{1.7}) invariant. This yields an overdetermined linear system of equations for the
infinitesimals $\xi_1$, $\xi_2$, $\xi_3$, $\phi_1$, $\phi_2$, $\phi_3$, $\phi_4$, $\phi_5$ and $\phi_6$. The
associated Lie algebra of infinitesimal symmetries is the set of vector fields of the form:
\begin{equation} X = \xi_1\frac {\partial}{\partial x} +  \xi_2\frac {\partial}{\partial y}+\xi_3\frac
{\partial}{\partial t}+\phi_1\frac {\partial}{\partial u}+\phi_2\frac {\partial}{\partial \omega}+\phi_3\frac
{\partial}{\partial m}+\phi_4\frac {\partial}{\partial \psi}+\phi_5\frac {\partial}{\partial
\chi}+\phi_6\frac {\partial}{\partial \lambda} . \label{6}\end{equation} By applying the  classical Lie
method \cite{stephani} to the (\ref{1.4})-(\ref{1.7}) system of PDEs , we obtain  a system of overdetermined
equations whose solutions are (We have used MACSYMA and MAPLE independently to handle the calculations):
\begin{eqnarray}
&&\xi_1 = k_1x+M_1\\&& \xi_2 =2k_1y+k_2\\&&
 \xi_3  = 3k_1t+k_3\end{eqnarray}
\begin{eqnarray}
 &&\phi_1  =u\left(M_2+\frac{dM_1}{dt}y-k_1\right)\label{2.13}\\&& \phi_2  =
\omega\left(-M_2-\frac{dM_1}{dt}y-k_1\right)\label{2.14}\\&&  \phi_3 =
M_3-\frac{1}{4}\frac{d^2M_1}{dt^2}y^2-\frac{1}{2}\frac{dM_2}{dt}y-k_1m \label{2.15}\\&&
\phi_4=\psi\left(\frac{1}{2}\frac{dM_2}{dt}+\frac{1}{2}\frac{dM_1}{dt}y+N\right)\label{2.16}\\&&
\phi_5=\chi\left(-\frac{1}{2}\frac{dM_2}{dt}-\frac{1}{2}\frac{dM_1}{dt}y+N\right)\label{2.17}\\&&
\phi_6=-k_1\lambda,\label{2.18}
\end{eqnarray}
where $k_1$, $k_2$ and $k_3$ are arbitrary constants; $M_i=M_i(t), (i=1..3)$ are arbitrary functions of $t$,
and $N=N(y,t)$ is a function of $y$ and $t$ that satisfies:
\begin{equation} N_t-\lambda N_y=0 .\label{2.19}\end{equation}

Having determined the infinitesimals   in (\ref{2.13})-(\ref{2.18}), the symmetry variables are found by
solving the corresponding characteristic equations
\begin{eqnarray}
&&\frac{dx}{k_1x+M_1}=\frac{dy}{2k_1y+k_2}=\frac{dt}{3k_1t+k_3}=\frac {-\,d\lambda}{k_1\lambda}=\nonumber\\&&
=\frac{du}{u\left(M_2+\frac{dM_1}{dt}y-k_1\right)}=\frac{-\,d\omega}{\omega\left(M_2+\frac{dM_1}{dt}y+k_1\right)}=\nonumber\\&&
=\frac{dm}{ M_3-\frac{1}{2}\frac{dM_2}{dt}y-\frac{1}{4}\frac{d^2M_1}{dt^2}y^2-k_1m }=\label{2.20}\\&&
=\frac{d\psi}{\psi\left(\frac{1}{2}M_2+\frac{1}{2}\frac{dM_1}{dt}y+N\right)}=
\frac{-\,d\chi}{\chi\left(\frac{1}{2}M_2+\frac{1}{2}\frac{dM_1}{dt}y-N\right)}.\nonumber
\end{eqnarray}
In  the next section we solve  (\ref{2.20}) for the different possibilities.

\section{Reductions}
There are four independent reductions, depending on the values of $k_1,k_2,k_3$:
\subsection{$k_1\neq 0$}

In this case, there is not restriction in setting $k_2=k_3=0$ because it only implies a trivial translation
in $y$ and $t$.

$\bullet$ By solving (\ref{2.20}), we have the reduced variables:
\begin{equation}z_1=\frac{x}{(3t)^{\frac{1}{3}}}-\frac{1}{k_1}\int \frac{M_1}{(3t)^{\frac{4}{3}}}dt,
\qquad z_2=\frac{y}{(3t)^{\frac{2}{3}}} . \label{3.1}\end{equation}

$\bullet$ The reduction of the spectral parameter is:
\begin{equation}
\lambda(y,t)=2(3t)^{-\frac{1}{3}}\Lambda(z_2) , \label{3.2}\end{equation} where (\ref{1.8}) yields the
following equation for $\Lambda(z_2)$:
\begin{equation}
2(z_2+\Lambda)\frac{d \Lambda}{dz_2}+\Lambda=0. \label{3.3}\end{equation} Therefore, \textbf{the reduced
linear problem is nonisospectral}.

$\bullet$ The reductions for the fields and eigenfunctions are:

 \begin{eqnarray}
&&u(x,y,t)= (3t)^{-\frac{1}{3}}e^{\displaystyle{2\,\Omega[t,z_2]}}\alpha(z_1,z_2)\label{3.4}\\
&&\omega(x,y,t)=(3t)^{-\frac{1}{3}}e^{{\displaystyle -2\,\Omega[t,z_2]}}\beta(z_1,z_2)\label{3.5}\\&&
m(x,y,t)=(3t)^{-\frac{1}{3}}\left(\gamma(z_1,z_2)+\Delta[t,z_2]\right)\label{3.6}\\&&
\psi(x,y,t)=e^{\displaystyle{\Omega[t,z_2]}}e^{\displaystyle {H[t,z_2]}}Q(z_1,z_2)\label{3.7}\\&&
\chi(x,y,t)=e^{{\displaystyle-\Omega[t,z_2]}} e^{\displaystyle
{H[t,z_2]}}R(z_1,z_2),\label{3.8}\end{eqnarray} where $\alpha(z_1,z_2),\, \beta(z_1,z_2),\,
\gamma(z_1,z_2),\, Q(z_1,z_2),\,R(z_1,z_2) $ are the reduced fields and eigenfunctions.

Furthermore  $\Omega[t,z_2]$, $H[t,z_2]$, $\Delta[t,z_2]$ are defined as:
\begin{eqnarray}
&&\Omega[t,z_2]=\frac{1}{2k_1}\left[\left(\frac{M_1}{(3t)^{\frac{1}{3}}}+\int \frac{M_1}{(3t)^{\frac{4}{3}}}dt\right)
z_2+\int \frac{M_2}{3t}dt\right]\label{3.9}\\
&&\Delta[t,z_2]=\frac{1}{2k_1}\left[\frac{M_1}{(3t)^{\frac{1}{3}}}+\int
\frac{M_1}{(3t)^{\frac{4}{3}}}dt-\frac{1}{2}\int (3t)^{\frac{2}{3}}\frac{dM_1}{dt}dt
\right]z_2^2+\nonumber\\&&\quad\quad+\frac{1}{2k_1}\left[-z_2M_2+2\int
\frac{M_3}{(3t)^{\frac{2}{3}}}dt\right]\label{3.10}\\&& H[t,z_2]=\frac{1}{k_1}\int \frac{\hat
N[t,z_2]}{3t}dt, \label{3.11}\end{eqnarray} where $$\hat N[t,z_2)]=N[t,y(t,z_2)].$$ Note that with this
definition of $\hat N[t,z_2]$, equation (\ref{2.19}) yields:
$$\frac {\partial \hat N}{\partial z_2}=\frac{1}{2}\left(\frac{3t}{z_2+\Lambda(z_2)}\right)\frac {\partial \hat N}{\partial t}.$$

$\bullet$ Substitution of the reductions in the $2+1$ spectral problem (\ref{1.4})-(\ref{1.7})  gives us  the
following $1+1$ Lax
 pair:

\begin{eqnarray}
&& Q_{z_1}+\Lambda Q+\alpha R=0 \label{3.12}\\
&&R_{z_1}-\Lambda R+\beta Q=0  \label{3.13} \\&& \nonumber 2\left(z_2+\Lambda\right)Q_{z_2}=
Q\left(\gamma_{z_2}+\alpha\beta_{z_1}-\beta\alpha_{z_1}+2\Lambda\alpha\beta-4\Lambda^3-\Lambda_{z_2}+z_1\Lambda
\right)+\\&& \quad\quad\quad
+R\left(-\alpha_{z_2}-\alpha_{z_1z_1}+2\beta\alpha^2-4\Lambda^2\alpha+2\Lambda\alpha_{z_1}+z_1\alpha
\right)\label{3.14}\\&& \nonumber 2\left(z_2+\Lambda\right)
R_{z_2}=R\left(-\gamma_{z_2}-\alpha\beta_{z_1}+\beta\alpha_{z_1}-2\Lambda\alpha\beta+4\Lambda^3-\Lambda_{z_2}-z_1\Lambda
\right)+\\&& \quad\quad\quad
+Q\left(\beta_{z_2}-\beta_{z_1z_1}+2\alpha\beta^2-4\Lambda^2\beta-2\Lambda\beta_{z_1}+z_1\beta\right).\label{3.15}
\end{eqnarray}

 $\bullet$ It is trivial to check that the compatibility condition of equations (\ref{3.12})-(\ref{3.15})  yields the
 system of \textbf{nonautonomous} equations:

\begin{eqnarray} &&\left(\gamma_{z_1}+\alpha\beta\right)_{z_2}=0\label{3.16}\\&&
 \alpha_{z_1z_1z_1}+\alpha_{z_1z_2}-
6\alpha\beta\alpha_{z_1}+2\alpha \gamma_{ z_2}-2z_2\alpha_{ z_2}-z_1\alpha_{z_1}-\alpha=0\label{3.17}\\&&
 \beta_{z_1z_1z_1}- \beta_{z_1z_2}-
6\alpha\beta \beta_{z_1}-2\beta\gamma_{z_2}-2z_2\beta_{z_2}-z_1\beta_{z_1}-\beta=0.\label{3.18}
\end{eqnarray}
It is not difficult to check that (\ref{3.16})-(\ref{3.18}) has the Painlev\'e property. This property and
the existence of the nonlinear associated problem (\ref{3.12})-(\ref{3.15}) prove that the system is
integrable although the spectral parameter is not a constant because it should satisfy the
\textbf{nonisospectral} condition (\ref{3.3}).

\subsection{$k_1=0,\quad k_2\neq 0,\quad k_3\neq 0$}
$\bullet$ The reduced variables provided by integration of (\ref{2.20}) are:
\begin{equation}z_1=\frac{k_2}{6k_3}x-\frac{k_2}{18 k_3^2}\int M_1dt,\qquad z_2=\left(\frac{k_2}{6k_3}\right)^2y-2\left(\frac{k_2}{6k_3}\right)^3t. \label{3.19}\end{equation}
$\bullet$ The  spectral parameter is reduced as:
\begin{equation}
\lambda(y,t)=\frac{k_2}{3k_3}\Lambda_0 , \label{3.20}\end{equation} where $\Lambda_0$ is a constant and
therefore, \textbf{the reduced linear problem is isospectral}.

$\bullet$ The  reductions for the fields are:

\begin{eqnarray}
&&u(x,y,t)=\frac{k_2}{6k_3}e^{\displaystyle{2\,\Omega[t,z_2]}}\alpha(z_1,z_2)\label{3.21}\\&&
\omega(x,y,t)=\frac{k_2}{6k_3}e^{{\displaystyle -2\,\Omega[t,z_2]}}\beta(z_1,z_2)\label{3.22}\\&&
m(x,y,t)=\frac{k_2}{6k_3}\gamma(z_1,z_2)+\frac{1}{3k_3}\Delta[t,z_2]\label{3.23}\\&&
\psi(x,y,t)=e^{\displaystyle{\Omega[t,z_2]}}e^{\displaystyle{H[t,z_2]}}Q(z_1,z_2)\label{3.24}\\&&
\chi(x,y,t)=e^{{\displaystyle-\Omega[t,z_2]}} e^{\displaystyle{H[t,z_2]}}R(z_1,z_2),\label{3.25}
\end{eqnarray} where the  functions $\Omega[t,z_2]$, $H[t,z_2]$ and $\Delta[t,z_2]$ are:

\begin{eqnarray} &&\Omega[t,z_2]=\frac{1}{6k_3}\left(\left[\left(\frac{6k_3}{k_2}\right)^2M_1\right]z_2+\frac{k_2}{3k_3}\int t\frac{dM_1}{dt}dt+\int M_2 dt\right)\label{3.26}\\&&
 \Delta[t,z_2]=\left[-\frac{1}{4}\left(\frac{6k_3}{k_2}\right)^4\frac{dM_1}{dt}\right]z_2^2+\left[-\frac{1}{2}\left(\frac{6k_3}{k_2}\right)^2M_2-\left(\frac{6k_3}{k_2}\right)\left(t\frac{dM_1}{dt}-M_1\right)\right]z_2\nonumber\\&&
\quad\quad\quad\quad +\left[\int M_3 dt-\left(\frac{k_2}{6k_3}\right)\int t\frac{dM_2}{dt}
-\left(\frac{k_2}{6k_3}\right)^2\int t^2\frac{d^2M_1}{dt^2}dt\right]\label{3.27}\\&&
 H[t,z_2]=\frac{1}{3k_3}\int \hat N[t,z_2] dt , \label{3.28}\end{eqnarray}
 where
 $$\hat N[t,z_2]=N[t,y(t,z_2)].$$
 Therefore,  (\ref{2.19}) yields
$$\frac {\partial \hat N}{\partial z_2}=\left(\frac{6k_3}{k_2}\right)^2\left(\frac{1}{1+\Lambda_0}\right)\frac {\partial \hat N}{\partial t}.$$

$\bullet$ The reduction of the Lax pair is:

\begin{eqnarray}
&&0= Q_{z_1}+\Lambda_0Q+\alpha R \label{3.29}\\
&&0=R_{z_1}-\Lambda_0 R+\beta \,Q  \label{3.30} \\
&&\nonumber 2\left(1+\Lambda_0\right)
Q_{z_2}=Q\left(\gamma_{z_2}+\alpha\,\beta_{z_1}-\beta\,\alpha_{z_1}+2\Lambda_0\,\alpha\,\beta
-4\Lambda_0^3\right)+\\&& \quad\quad\quad
+R\left(-\alpha_{z_2}-\alpha_{z_1z_1}+2\beta\,\alpha^2-4\Lambda_0^2\,\alpha+2\Lambda_0\,\alpha_{z_1}
\right)\label{3.31}\\
&&\nonumber 2\left(1+\Lambda_0\right) R_{z_2}=R\left(-\gamma_{z_2}-\alpha\,\beta_{z_1}+\beta\,\alpha_{z_1}
-2\Lambda_0\,\alpha\,\beta+4\Lambda_0^3\right)+\\&& \quad\quad\quad
+Q\left(\beta_{z_2}-\beta_{z_1z_1}+2\alpha\,\beta^2
-4\Lambda_0^2\,\beta-2\Lambda_0\,\beta_{z_1}\right).\label{3.32}
\end{eqnarray}
$\bullet$ The compatibility condition of this Lax pair gives us the system:

\begin{eqnarray} &&\left(\gamma_{z_1}+\alpha\beta\right)_{z_2}=0\label{3.33}\\&&
 \alpha_{z_1z_1z_1}+\alpha_{z_1z_2}-
6\alpha\beta\alpha_{z_1}+2\alpha \gamma_{ z_2}-2\alpha_{ z_2}=0\label{3.34}\\&&
 \beta_{z_1z_1z_1}- \beta_{z_1z_2}-
6\alpha\beta \beta_{z_1}-2\beta\gamma_{z_2}-2\beta_{z_2}=0.\label{3.35}
\end{eqnarray}

\subsection{$k_1=0,\quad k_2\neq 0,\quad k_3= 0$}
$\bullet$ In this case, (\ref{2.20}) indicates that $t$ is one of the reduced variables. The other reduced
variable is $z_1$ defined by: \begin{equation} z_1=x-\frac{M_1 }{k_2}y. \label{3.36}\end{equation} $\bullet$
For the spectral parameter we have the reduction:
\begin{equation}
\lambda(y,t)=2\Lambda_0 , \label{3.37}\end{equation} where (\ref{3.37}) implies that $\Lambda_0$ is a
constant, and therefore the reduced problem is isospectral.
 $\bullet$ The reduction for the fields is:

\begin{eqnarray}
&&u(x,y,t)=e^{\displaystyle{2\,\Omega(y,t)}}\alpha(z_1,t)\label{3.38}\\&& \omega(x,y,t)=e^{{\displaystyle
-2\,\Omega(y,t)}}\beta(z_1,t)\label{3.39}\\&& m(x,y,t)=\gamma(z_1,t)+\frac{1}{k_2}\Delta(y,t)\label{3.40}\\&&
\psi(x,y,t)=e^{\displaystyle{\Omega(y,t)}}e^{\displaystyle{H(y,t)}}Q(z_1,t)\label{3.41}\\&&
\chi(x,y,t)=e^{{\displaystyle-\Omega(y,t)}} e^{\displaystyle{H(y,t)}}R(z_1,t),\label{3.42}\end{eqnarray}

 where $\Omega(y,t)$, $H(y,t)$ and $\Delta(y,t)$ are:

\begin{eqnarray} &&\Omega(y,t)=\frac{1}{2k_2}\left(M_2\,y+\frac{1}{2}\frac{dM_1}{dt}y^2\right)\label{3.43}\\&&
 \Delta(y,t)=\frac{1}{k_2}\left(M_3\,y-\frac{1}{4}\frac{dM_2}{dt}y^2-\frac{1}{12}\frac{d^2M_1}{dt^2}y^3 \right)\label{3.44}\\&&
 H(y,t)=\frac{1}{k_2}\int\ N(y,t)\, dy . \label{3.45}\end{eqnarray}

$\bullet$ The reduced spectral problem is: \begin{eqnarray} &&0= Q_{z_1}+\lambda_0\,Q+\alpha R
\label{3.46}\\&& 0=R_{z_1}-\lambda_0R+\beta Q \label{3.47}
\\&& \nonumber
Q_{t}=Q\left(\beta\alpha_{z_1}-\alpha\beta_{z_1}-2\Lambda_0\,\alpha\beta+4\Lambda_0^3+\frac{1}{k_2}\left[-M_3+\Lambda_0\,M_2+2\Lambda_0^2\,M_1+M_1\gamma_{z_1}\right]\right)+\\&&
\quad+R\left(\alpha_{z_1z_1}-2\,\beta\,\alpha^2+4\Lambda_0^2\alpha
-2\Lambda_0\,\alpha_{z_1}+\frac{1}{k_2}\left[2\Lambda_0\,M_1\alpha+M_2\,\alpha-M_1\,\alpha_{z_1}\right]\right)\label{3.48}\\&&
\nonumber
R_{t}=R\left(-\beta\alpha_{z_1}+\alpha\beta_{z_1}+2\Lambda_0\,\alpha\beta-4\Lambda_0^3+\frac{1}{k_2}\left[
M_3-\Lambda_0\,M_2-2\Lambda_0^2\,M_1-M_1\gamma_{z_1}\right]\right)+\\&&
\quad+Q\left(\beta_{z_1z_1}-2\,\alpha\,\beta^2+4\Lambda_0^2\beta
+2\Lambda_0\,\beta_{z_1}+\frac{1}{k_2}\left[2\Lambda_0\,M_1\beta+M_2\,\beta+M_1\,\beta_{z_1}\right]\right),\label{3.49}
\end{eqnarray}
where $\Lambda_0$ is the spectral parameter.

$\bullet$ The compatibility condition of (\ref{3.46})-(\ref{3.49}) is the following system of  PDEs:

 \begin{eqnarray} &&\left(\gamma_{z_1}+\alpha\,\beta\right)_{z_1}=0\label{3.50}\\&&
 \alpha_{z_1z_1z_1}+\alpha_{t}-
6\alpha\,\beta\,\alpha_{z_1}+\frac{M_2\,\alpha_{z_1}+2M_3\,\alpha-M_1\,\left(\alpha_{z_1z_1}+2\alpha\,\gamma_{z_1}\right)}{k_2}=0
\label{3.51}\\&&
 \beta_{z_1z_1z_1}+\beta_{t}
-6\alpha\,\beta\,\beta_{z_1}+\frac{M_2\,\beta_{z_1}-2M_3\,\beta +
M_1\,\left(\beta_{z_1z_1}+2\,\beta\,\gamma_{z_1}\right)}{k_2}=0 ,  \label{3.52}
\end{eqnarray}
which are the reduction of (\ref{1.1})-(\ref{1.3}). This case includes for $M_2=M_3=0$ the Hirota equation of
reference \cite{Hirota}.

\subsection{$k_1=0,\quad k_2= 0,\quad k_3\neq 0 $}
$\bullet$ $y$ is now one of the reduced variables. The other reduced variable is $z_1$, defined by:
\begin{equation} z_1=x-\frac{\int M_1(t)dt}{3k_3}. \label{3.53}\end{equation}
$\bullet$  The spectral parameter is reduced as follows:
 \begin{equation}\Lambda(z_1,y)=2\Lambda_0 , \label{3.54}\end{equation}
 with $\lambda_0$ an arbitrary constant.

$\bullet$ The reduction for the fields is:
\begin{eqnarray}
&&u(x,y,t)=e^{\displaystyle{2\,\Omega(y,t)}}\alpha(z_1,y)\label{3.55}\\&& \omega(x,y,t)=e^{{\displaystyle
-2\,\Omega(y,t)}}\beta(z_1,y)\label{3.56}\\&& m(x,y,t)=\gamma(z_1,y)+\Delta(y,t)\label{3.57}\\&&
\psi(x,y,t)=e^{\displaystyle{\Omega(y,t)}}e^{\displaystyle{H(y,t)}}Q(z_1,y)\label{3.58}\\&&
\chi(x,y,t)=e^{{\displaystyle-\Omega(y,t)}} e^{\displaystyle{H(y,t)}}R(z_1,y,)\label{3.59}\end{eqnarray}

 where $\Omega(y,t)$, $H(y,t)$ and $\Delta(y,t)$ are:

\begin{eqnarray} &&\Omega(y,t)=\frac{1}{6k_3}\left(M_1\,y+\int M_2(t)dt\right)\label{3.60}\\&&
 \Delta(y,t)=\frac{1}{3k_3}\left(\int M_3(t)dt-\frac{M_2}{2}\,y-\frac{1}{4}\frac{dM_1}{dt}\,y^2 \right)\label{3.61}\\&&
 H(y,t)=\frac{1}{3k_3}\int N(y,t) dt. \label{3.62}\end{eqnarray}

$\bullet$ The reduction of the Lax pair yields:
\begin{eqnarray}
&& 0= Q_{z_1}+\Lambda_0\,Q+\alpha \,R \label{3.63}\\&& 0=R_{z_1}-\Lambda_0\,R+\beta \,Q \label{3.64}
\\&& \nonumber
2\Lambda_0\,Q_{y}=Q\left(\gamma_y+\alpha\,\beta_{z_1}-\beta\,\alpha_{z_1}+2\Lambda_0\,\alpha\,\beta-4\Lambda_0^3
\right)+\\&&
\quad+R\left(-\alpha_{z_1z_1}-\alpha_y+2\alpha^2\,\beta+2\Lambda_0\,\alpha_{z_1}-4\Lambda_0^2\,\alpha\right)\label{3.65}\\&&
\nonumber
2\Lambda_0R_{y}=R\left(-\gamma_y-\alpha\,\beta_{z_1}+\beta\,\alpha_{z_1}-2\Lambda_0\,\alpha\,\beta+4\Lambda_0^3
\right)+\\&& \quad
+Q\left(-\beta_{z_1z_1}+\beta_y+2\alpha\,\beta^2-2\Lambda_0\,\beta_{z_1}-4\Lambda_0^2\,\beta\right) ,
\label{3.66}
\end{eqnarray}
whose  compatibility condition of (\ref{3.63})-(\ref{3.66}) is the system of equations:

\begin{eqnarray} &&\left(\gamma_{z_1}+\alpha\beta\right)_{z_2}=0\label{3.67}\\&&
 \alpha_{z_1z_1z_1}+\alpha_{z_1y}-
6\alpha\beta\alpha_{z_1}+2\alpha\gamma_y=0 \label{3.68}\\&&
 \beta_{z_1z_1z_1}-\beta_{z_1y}
-6\alpha\beta\beta_{z_1}-2\beta\gamma_y=0 . \label{3.69}
\end{eqnarray}

\section{Conclusions}

A spectral problem in $2+1$ dimensions is presented. The compatibility conditions of this Lax pair yields a
$2+1$ system that was introduced in \cite{Maccari98}. An important fact is that the spectral parameter is
nonisospectral.

If we wish to know the $1+1$ reductions of the spectral problem, it is specially important to establish how
the spectral parameter reduces. One possibility is to identify the classical Lie symmetries of the Lax pair
instead of those of the system of PDEs. We have identify these symmetries \textbf{by considering the spectral
parameter as an additional field}. This means that we obtain symmetries that are symmetries of the fields $m,
u,\omega$, the eigenfunctions $\psi, \chi$ and the spectral parameter $\lambda$. The symmetries that we have
obtained include three arbitrary constants and several arbitrary functions.

We attempt to go from $2+1$ to $1+1$ dimensions by using the reductions arising from the former Classical Lie
symmetries.
  Four possible reductions arise from the Classical Symmetries. They yield highly nontrivial systems of nonautonomous
PDEs in $1+1$ dimensions as well as their associated spectral problems, with spectral parameters that are
obtained from reductions of the function $\lambda(y,t)$. Particularly interesting is  case \textit{3.1},
where the spectral parameter is not a constant even in the $1+1$ reduction.

Obviously, classical Lie symmetries are not the only ones that can be identified. Other symmetries such
as nonclassical or potential symmetries can be studied in the future. Nevertheless the purpose of this paper
is not an exhaustive study of the symmetries of (\ref{1.1})-(\ref{1.3}) but to proof that the study of the
symmetries of the Lax pair and the spectral parameter provides much more interesting information that the
symmetries of the system of PDE. This information is specially relevant when we obtain the similarity
reductions that provides  the reduced spectral problem that yields to the reduced system and the reduction of
the spectral parameter.

\section*{Acknowledgements}
This research has been supported in part by the DGICYT under project  FIS2009-07880 and JCyL under contract
GR224.


\begin{thebibliography}{99}
\bibitem{ars} Ablowitz M~J, Ramani A and Segur H, Nonlinear evolution equations and ordinary differential equations of Painlev\'e type, {\it
Lett. Nuov. Cim.} {\bf 23} (1978) 333--338
\bibitem{bluco}
Bluman G~W and Cole J~D, \textit{Similarity Methods for Differential Equations}, Springer Verlag, (1974)
\bibitem{calogero} Calogero F and Degasperis A, Nonlinear evolution equations solvable by the inverse spectral
transform-I, {\it Nuovo Cimento Soc. Ital. Fis. B} \textbf{32} (1976) 201--242
\bibitem{estevez01} Est\'evez P~G, A nonisospectral problem in $(2+1)$ dimensions derived from KP, {\it Inverse Problems} {\bf 17} (2001)
1043-1052
\bibitem{ep01} Est\'evez P~G and Prada J, Singular Manifold Method for an Equation in
2 + 1 Dimensions, {\it Journal of Nonlinear Mathematical Physics} {\bf 12} (2005) 266--279
\bibitem{egp05} Est\'evez  P~G, Gandarias M~L  and Prada J, Symmetry reductions of a $2 +1$ Lax pair, {\it Phys. Lett. A} {\bf 343} (2005)
40--47
\bibitem{Hirota} Hirota R, Exact envelope-soliton solutions of a nonlinear wave equation, {\it J. Math. Phys.}
\textbf{14} (1973) 805--809
\bibitem{Maccari98}
Maccari A, A generalized Hirota equation in $2+1$ dimensions, {\it J. Math. Phys.} {\bf 39} (1998) 6547--6551
\bibitem{legar96}Legar\'e M, Symmetry Reductions of the Lax Pair of the Four-Dimensional Euclidean Self-Dual Yang-Mills
Equations, {\it Journal of Nonlinear Mathematical Physics} {\bf 3} (1996) 266--285
\bibitem{olver}
Olver P~J, \textit{Applications of Lie Groups to Differential Equations}, Springer Verlag, (1999)
\bibitem{pp} Painlev\'e P, Sur les equations differentielles du second ordre et d'ordre superieur dont l'integrale
gen´erale est uniforme, {\it Acta Math.}\textbf{ 25} (1902) 1–-85
\bibitem{stephani}
Stephani H, \textit{Differential equations. Their solutions using symmetries}, edited by M. Mac Callum,
Cambridge University Press, (1989)
\bibitem{Weiss} Weiss J, The Painlev\'e property for partial differential equations II: B\"acklund transformation, Lax pairs, and the Schwarzian derivative, {\it J. Math. Phys.} {\bf 24} (1983) 1405--1413
\end{thebibliography}
\end{document}